\documentclass[apj]{emulateapj}
\usepackage{amsmath}
\usepackage{natbib}
\usepackage{color}
\usepackage{url}

\newcommand{\clustername}{SDSS~J1531+3414}
\newcommand{\clusterz}{0.335}
\newcommand{\zarcA}{1.096}

\newcommand{\hst}{{\it HST}}

\newcommand{\msun}{M$_{\odot}$}

\newcommand{\implanerms}{$0.23$}

\newcommand{\Px}{$-1.24 _{-0.16}^{+0.13}$}
\newcommand{\Py}{$-0.73 ^{+0.22}_{-0.15}$}
\newcommand{\Pe}{$0.11 ^{+0.04}_{-0.01}$}
\newcommand{\Ptheta}{$87.6 ^{+4.8}_{-10.6}$}
\newcommand{\Prc}{$31.9 ^{+4.9}_{-7.7}$}
\newcommand{\Psigma}{$791 ^{+ 34}_{- 55}$}
\newcommand{\PxA}{$49.9 _{-20.5}^{+3.1}$}
\newcommand{\PyA}{$-45.6 ^{+25.1}_{-4.0}$}
\newcommand{\PeA}{$0.93 ^{+0.02}_{-0.22}$}
\newcommand{\PthetaA}{$57.0 ^{+12.4}_{-4.9}$}
\newcommand{\PsigmaA}{$631 ^{+ 82}_{-152}$}
\newcommand{\PredshiftB}{$2.01 ^{+0.39}_{-0.13}$}
\newcommand{\PredshiftC}{$1.55 \pm0.05$}
\newcommand{\mass}{$5.9 ^{+ 0.2}_{- 0.3}\times 10^{13}$\msun}
\newcommand{\Paxy}{14.6}
\newcommand{\Pbxy}{-1.2}
\newcommand{\Pazsig}{1159.7}
\newcommand{\Pbzsig}{-180.2}
\newcommand{\Pazmass}{77.4}
\newcommand{\Pbzmass}{-9.1}
\newcommand{\thetaE}{$11.2 \pm{0.1}\arcsec$}
\newcommand{\thetaEB}{$16.1 ^{+0.6}_{-0.5}\arcsec$}
\newcommand{\sigmacorr}{$1.01$}
\newcommand{\Prcorercut}{$0.02$}
\newcommand{\Psigmacorr}{$802 ^{+ 34}_{- 55}$}

\shorttitle{The lens model of \clustername}
\slugcomment{Submitted to ApJ: draft date \today}
\shortauthors{Sharon et al.}

\begin{document}
\title{The mass distribution of the strong lensing cluster \clustername\altaffilmark{*}}
\altaffiltext{*}{Based on observations made with the NASA/ESA 
  {\it Hubble Space Telescope}, obtained at the Space Telescope Science
  Institute, which is operated by the Association of Universities for
  Research in Astronomy, Inc., under NASA contract NAS 5-26555. These
  observations are associated with program GO-12267}
\author{Keren Sharon\altaffilmark{1}, 
{Michael D. Gladders\altaffilmark{2,3}}, 
{Jane R. Rigby\altaffilmark{4}}, 
{Eva Wuyts\altaffilmark{5}},
{Matthew B. Bayliss\altaffilmark{6,7}},
{Traci L.  Johnson\altaffilmark{1}},
{Michael K. Florian\altaffilmark{3}},
H{\aa}kon Dahle\altaffilmark{8}} 

\email{kerens@umich.edu} 

\altaffiltext{1}{Department of Astronomy, University of Michigan, 500 Church Street, Ann Arbor, MI 48109, USA} 
\altaffiltext{2}{Kavli Institute for Cosmological Physics, University of Chicago, 5640 South Ellis Avenue, Chicago, IL 60637, USA.}
\altaffiltext{3}{Department of Astronomy and Astrophysics, University of Chicago, 5640 South Ellis Avenue, Chicago, IL 60637, USA}
\altaffiltext{4}{Observational Cosmology Lab, NASA Goddard Space Flight Center, Greenbelt MD 20771}
\altaffiltext{5} {Max-Planck-Institut f{\"u}r extraterrestrische Physik, Giessenbachstr. 1, D-85741 Garching, Germany} 
\altaffiltext{6} {Department of Physics, Harvard University, 17 Oxford Street, Cambridge, MA 02138} 
\altaffiltext{7} {Harvard-Smithsonian Center for Astrophysics, 60 Garden Street, Cambridge, MA 02138, USA} 
\altaffiltext{8} {Institute of Theoretical Astrophysics, University of Oslo, P. O. Box 1029, Blindern, N-0315 Oslo, Norway}

\begin{abstract}
We present the mass distribution at the core of \clustername, a
strong-lensing cluster at $z=\clusterz$. We find that the mass
distribution is well described by two cluster-scale halos with a
contribution from cluster-member galaxies. New \hst~ observations of
\clustername~ reveal a signature of ongoing star formation associated
with the two central galaxies at the core of the cluster, in the form
of a chain of star forming regions at the center of the cluster. Using
the lens model presented here, we place upper limits on the
contribution of a possible lensed image to the flux at the central
region, and rule out that this emission is coming from a background
source.

\end{abstract}


\section{Introduction}
The number of galaxy clusters that are known to be strong lenses has
increased dramatically over the last decade, from a handful of
massive clusters, to hundreds of systems with masses ranging from
group-scale to super-massive nodes on the cosmic web. This progress is
made possible thanks to dedicated surveys that utilize wide field
imaging, cluster finding algorithms, and extensive followup. One such
survey is the Sloan Giant Arcs Survey (SGAS, Gladders et al. in prep;
Bayliss et al. 2011; Hennawi et al. 2008). In short, galaxy
clusters were optically
selected from the Sloan Digital Sky Survey (SDSS, York et al. 2000)
data, based on the red sequence technique (Gladders \& Yee 2000). Each
galaxy cluster field was then rendered to create a color image with
scaling parameters selected to enhance low signal-to-noise features,
and inspected by astronomers for evidence of strong lensing. 
Candidate lenses were imaged with larger telescopes, and all the high
confidence lensed galaxies (arcs) were followed up
spectroscopically. The multi-wavelength SGAS followup
includes spectroscopy of all the bright arcs (Bayliss et al. 2011), and for different
subsamples of the clusters, we have Subaru imaging for weak
lensing (Oguri et al. 2012), Spitzer imaging, UV imaging
(Bayliss 2012), radio observations 
of the Sunyaev-Zel'dovich (SZ) Effect (Gralla et al. 2011), 
detailed, medium
resolution and IFU spectroscopy (Wuyts et al. 2012a,b; Rigby et al. 2014; Bayliss
et al. 2013; Wuyts et al. 2014), and \hst~ imaging
(Bayliss et al. 2013). This multi-wavelength approach enables the
study of both the lensed background galaxies, some of which are
highly magnified (Wuyts et al. 2012a, Rigby et al. 2014, Koester et al,
Bayliss et al. 2010; Gladders et al. 2013; Dahle et al. 2013), and the lensing clusters
themselves (Gralla et al 2011; Bayliss et al. 2011, 2014; Blanchard et
al. 2013). 

High resolution \hst~ imaging as part
of \hst~ Cycle 20 program GO$-$13003 (PI: Gladders) was obtained for
37 SGAS cluster fields, in
order to study the lensed sources, and answer fundamental questions
about the galaxy population at the 
peak of the star formation history of the Universe. The main
scientific goal of this program is to determine the
morphology of star formation and the fundamental sizes of star
forming regions in galaxies at that epoch.
The combination of
lensing magnification by galaxy clusters and the \hst~ resolution can
uniquely access individual star forming regions in these
distant galaxies, and probe scales down to 100 pc. Typical SGAS arcs, by selection, are bright enough for
ground-based followup and detailed spectroscopy to explore the
physical conditions in these galaxies. On the other hand, they are not
necessarily intrinsically luminous, and are thus more representative
of their population than bright unlensed field galaxies. 

The study of the background source, as well as the mass distribution
at the core of the galaxy cluster,
relies on the availability of a robust lens model. The lensing
magnification and its uncertainty are needed in order to convert
measured quantities such as luminosity, star formation rate, stellar
mass and sizes to their unlensed values. Moreover, images of lensed galaxies can
be highly distorted, especially when they form giant arcs. The mass
model allows us to compute an image of the source galaxy, by
ray-tracing through the lensing equation (e.g., Sharon et al. 2012). 

\clustername~ was discovered as part of SGAS and first presented in Hennawi
et al. (2008) as a definite lensing cluster, based on ground-based imaging
detection of multiple prominent arcs. The cluster redshift, $z=0.335$,
was determined from SDSS spectroscopy of three galaxies, including the
brightest cluster galaxy (BCG). Followup spectroscopic observations
were executed in 2008 with 
the Gemini Multi-Object Spectrograph (GMOS; Hook et al. 2004),
yielding redshifts of eight cluster galaxies and three background sources
(Bayliss et al. 2011). Bayliss et al. also measured a velocity
dispersion of $998^{+120}_{-194}$ km s$^{-1}$ for this cluster, indicating that it
is moderately massive. 

Gralla et al. (2011) report an SZ mass and lensing analysis of
\clustername~ as part of a study of the mass-concentration relation in 10
lensing-selected clusters. 
They measured M$_{500}=1.7^{+0.4}_{-0.3}h^{-1}_{0.73}\times
10^{14}$ \msun, where $r_{500}=0.8$ Mpc, 
using an SZ-weak lensing scaling relation from Marrone et
al. (2012). They report an Einstein radius of $R_E=12.3 \arcsec$ for a
source at $z=1.096$ with an estimated
uncertainty of 5\%. 
Oguri et al. (2012) used Subaru imaging in a combined weak and
strong lensing analysis of 28 SGAS clusters, including
\clustername. They measured an 
Einstein radius of  $R_E = 11.7\pm1.2 \arcsec$ for a
source at $z=1.096$ and a weak lensing mass $M_{vir} = 5.13^{+1.33}_{-1.19}\times 10^{14}
h^{-1}_{0.702}$ \msun, in line with the velocity dispersion measurement,
within errors.  In both Gralla et al. (2011) and Oguri et al. (2012), the strong
lensing analysis was based on ground-based imaging, which led to 
 misinterpretation of some of the lensed features due to the poor
resolution.

Our recent \hst~ imaging data of \clustername~ revealed that what was
suspected to be light from images of radial arcs close to the central galaxies,
is in fact emission from ongoing star formation in the form of star
clusters forming a ``beads on
a string'' morphology. We report on this unusual galaxy-cluster star formation in a
companion paper, Tremblay et al. (2014).

In this paper, we present the mass distribution in the inner 100 kpc
of \clustername~ from detailed strong lensing model,
based on the new \hst~ imaging. We describe the \hst~  data in
\S~\ref{s.data}; the lensing analysis, including revised identification of
lensed galaxies and their redshifts, the lens modeling process, and the derived
model results are
described in \S~\ref{s.lensing}; finally, we discuss implications for
the star formation detected in the core
of \clustername~ in \S~\ref{s.core}.
Throughout the paper we assume a flat cosmology with $\Omega_{\Lambda}
= 0.73$, $\Omega_{m} =0.27$, and $H_0 = 73$  km s$^{-1}$
Mpc$^{-1}$. In this cosmology, $1\arcsec$ corresponds to 4.64 kpc at
the cluster redshift, $z=0.335$.
Magnitudes are reported in the AB system.

\section{Data}\label{s.data}
\clustername~ was observed as part of \hst~ Cycle~20 program GO$-$13003
(PI: Gladders) during three orbits on UT 2013 May 06. The observation was
restricted to the WFC3 cameras in order to reduce overheads, and split between
the IR and UVIS detectors. The filters were selected to
provide good sampling of the spectral energy distribution of the main
lensed galaxy, at $z=\zarcA$: 
in the UVIS channel, it was observed in three filters, F390W (2256 s),  F606W
(1440 s), and F814W (1964 s); and in the IR channel, it was observed with the
F160W filter (912 s). 
Each image was executed in four sub exposures with sub-pixel dithers
in order to fill the chip gap and to remove cosmic rays, hot or bad
pixels, and other artifacts.  
Individual images were combined onto a common grid using AstroDrizzle
(Gonzaga et al. 2012) with a pixel scale of $0\farcs03$ pixel$^{-1}$, and drop size of
0.5 for the IR filters and 0.8 for the UVIS filters. These values were
selected to provide the best sampling of the point spread function in each filter.
Images taken with the WFC3 camera are known to have several artifacts
that degrade the data quality. In the IR channel, circular areas with
lowered sensitivity are referred to as ``IR Blobs'' in the WFC3 Data
Handbook (Rajan et al. 2010).
We used a custom algorithm to remove the IR blobs from all the
images, by modeling these blobs in a superflat image created from all
the IR imaging in our program, and flatfielding these artifacts
out. This procedure was applied to each IR image prior to
drizzling. 
In the UVIS channel, the declining charge transfer efficiency (CTE) of
the detector can cause large flux losses and increase the level of
correlated noise in the image. 
To mitigate the CTE losses, the bluest UVIS data were taken with post-flash,
to increase the image background and ensure that faint sources have
high enough counts (see WFC3 Data Handbook, Raja et al. 2010). 
Post-observation image corrections were applied to individual
exposures using the Pixel-based Empirical CTE Correction
Software\footnote{\url{http://www.stsci.edu/hst/wfc3/ins\_performance/CTE/}}
provided by STScI. In the final reduced data, the limiting magnitudes for $5\sigma$
detection are m=26.5 mag in the F390W filter, and  m=26.1 mag in F606W,  F814W
and F160W.

\section{Lensing Analysis}\label{s.lensing}
\subsection{Identification of Lensed Images}\label{s.images}
The main lensed features in \clustername~ are detectable in ground-based
imaging, and were targeted for spectroscopy prior to the \hst~
observation, as part of a spectroscopic campaign to
secure redshifts of background sources and cluster galaxies in a large
sample of SGAS galaxies. These spectroscopic observations and analyses are described in Bayliss et
al. (2011), and we refer the reader to that publication for further details.  
For consistency, we mention the image IDs of Bayliss et al. (2011).

The lensing system of \clustername~ consists of three confirmed strongly-lensed
galaxies. We describe each of these systems below.

\begin{figure*}
\centering
\includegraphics[scale=0.46]{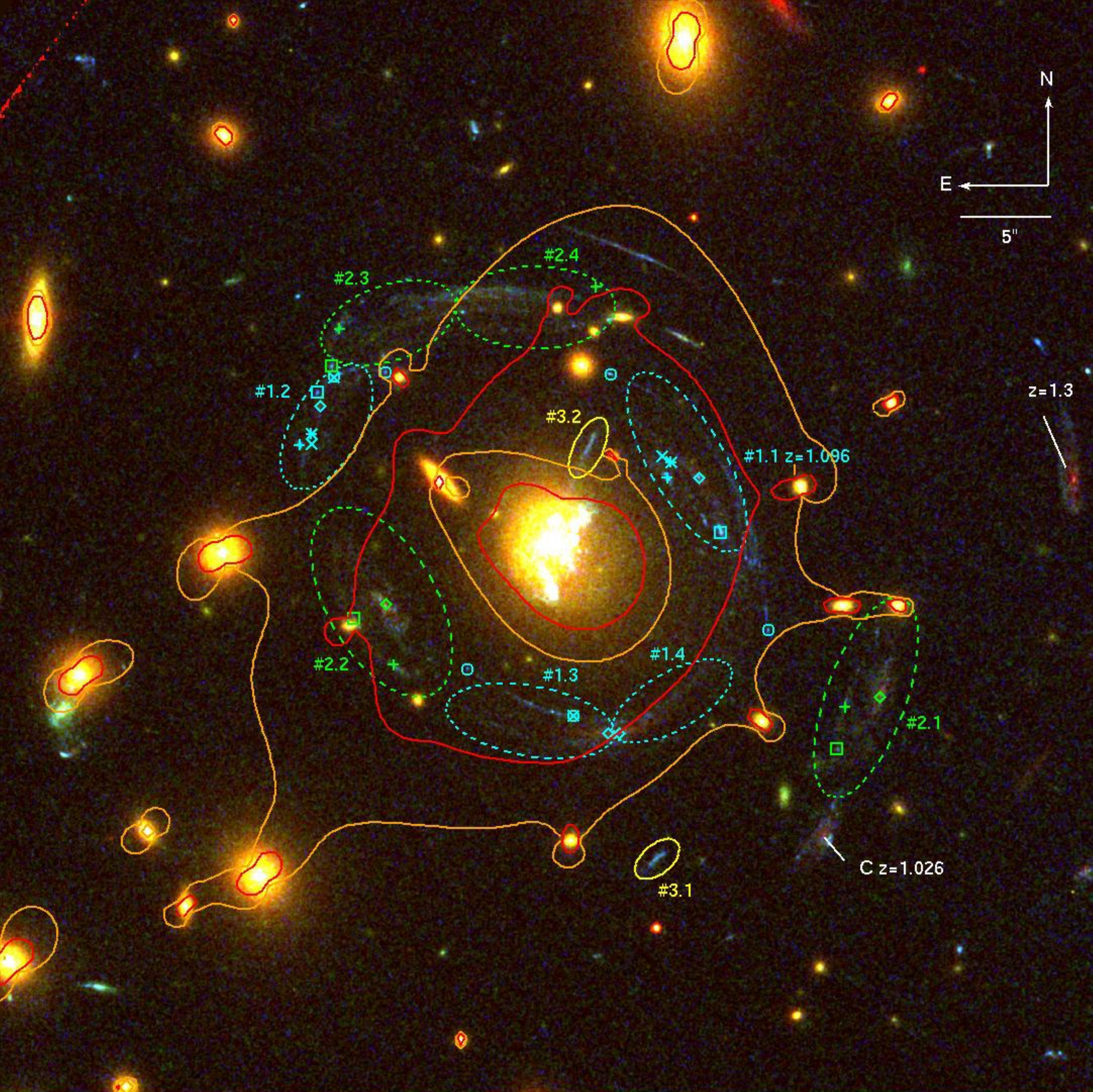}
\caption{{Color composite image of the core of the strong lensing cluster \clustername~ 
using {\it HST}/WFC3 F160W (red) ,F606W (green), F390W
(blue). The critical curves of the best-fit lens model are plotted in
red for $z=1.096$ and in orange for $z=2.01$. Multiply-lensed galaxies are
marked with ellipses, and their IDs and known redshifts are labeled. 
We mark features in these galaxies that were used as
constraints with symbols, to guide the eye to the matching features.
We also indicate other objects in the background of the clusters for
which we measured spectroscopic redshifts in Bayliss et al. (2011). At
the cluster redshift, $z=0.335$, $1\arcsec$ corresponds to 4.64 kpc. 
}}
\label{fig.lensmodel}
\end{figure*}

System \#1 is multiply imaged into five images (Figure~\ref{fig.lensmodel}). Images~1.1 and 1.2 are complete
images of the background galaxy, and 1.3 and 1.4 are partial images
forming a merging pair south of the cluster core. A fifth image is
predicted by the lens model to lie close to the central cluster
galaxies, but is not uniquely identified. Several emission knots in images 1.1, 1.2, 1.3 and 1.4 were targeted for
spectroscopy by Bayliss et al. (2011; A1-A4 in their Figure 5) and
show identical features, including strong emission
line from O~II,
which confirms the lensing interpretation.  The source is
spectroscopically confirmed to be at z=\zarcA~ (Bayliss et al. 2011). 

System~\#2 is composed of five images as well. Similarly to system~\#1, images 2.1 and 2.2 are
complete images of the background galaxy, and image 2.3 and 2.4 form a
merging pair at the north of the cluster center. A fifth image is
predicted behind the central galaxies, and is not identified. The redshift of Source~\#2 is not
spectroscopically determined. Bayliss et al. (2011) targeted images 2.1, 2.2,
2.3 for spectroscopy: some of these spectra show a blue continuum
with no features, providing a lower limit of $z>1.49$, and the other
spectra were not useful for measuring redshifts due to contamination
of one or both nod-and-shuffle traces by chance collision with other
galaxies and stars.  
We note that in Bayliss et al. (2011), image 2.3 is
labeled as B1, and due to a typographical error it was erroneously
reported to be at $z= 1.3$. 

Source~\#3 is multiply imaged into a tangential arc (3.1) and radial
arc (3.2). A
spectrum of the radial arc was obtained by Bayliss et al. (2011) and
shows low signal-to-noise blue continuum with no features, which
supports a lower limit of  $z>1.49$. Image 3.1 was not targeted for
spectroscopy.

Images A5, A6, B2, and C1 in Bayliss et al. (2011) are found to be not
multiply-imaged and are thus not used as constraints in this work.

\begin{deluxetable}{lcccc} 
 \tablecolumns{5} 
\tablecaption{Model constraints  \label{table.arcs}} 
\tablehead{\colhead{Source}   & 
            \colhead{ID }   & 
            \colhead{RA [$^o$]}     & 
            \colhead{Dec  [$^o$]}    & 
            \colhead{$z$}      } 
\startdata 
\#1  & 11.1 & 232.798460  & 34.242414 &   1.096  \\
   & 11.2 & 232.791470  & 34.241359 &     \\
   & 11.3 & 232.792900  & 34.237469 &     \\
   & 11.4 & 232.793120  & 34.237454 &     \\
   & 12.1 & 232.798830  & 34.241860 &     \\
   & 12.2 & 232.792040  & 34.241349 &     \\
   & 13.1 & 232.798630  & 34.241870 &     \\
   & 13.2 & 232.792140  & 34.241676 &     \\
   & 14.1 & 232.798610  & 34.242029 &     \\
   & 14.2 & 232.791970  & 34.241605 &     \\
   & 15.1 & 232.798500  & 34.242660 &     \\
   & 15.2 & 232.791060  & 34.240519 &     \\
   & 16.1 & 232.798210  & 34.242889 &     \\
   & 16.2 & 232.793790  & 34.237723 &     \\
   & 17.1 & 232.797250  & 34.242966 &     \\
   & 17.2 & 232.793110  & 34.242933 &     \\
   & 17.3 & 232.795740  & 34.238452 &     \\
   & 17.4 & 232.790180  & 34.239033 &     \\
\hline 
\#2   & 21.1 & 232.788120  & 34.238060 &  $>1.49$   \\
   & 21.2 & 232.797260  & 34.239392 &     \\
   & 22.1 & 232.788770  & 34.237850 &     \\
   & 22.2 & 232.797100  & 34.238497 &     \\
   & 22.3 & 232.798090  & 34.243633 &     \\
   & 22.4 & 232.793370  & 34.244258 &     \\
   & 23.1 & 232.788910  & 34.237212 &     \\
   & 23.2 & 232.797800  & 34.239135 &     \\
   & 23.3 & 232.798250  & 34.243064 &     \\
\hline 
 \#3  & 30.1 & 232.793550  & 34.241752 &   $>1.49$  \\
   & 30.2 & 232.792240  & 34.235538 &     
\enddata 
 \tablecomments{Positions of the lensed features that were used as
   constraints in the lens model. The redshifts and redshift limits
   are from Bayliss et al. (2011).}
\end{deluxetable}

\begin{figure*}
\centering
\includegraphics[scale=0.7]{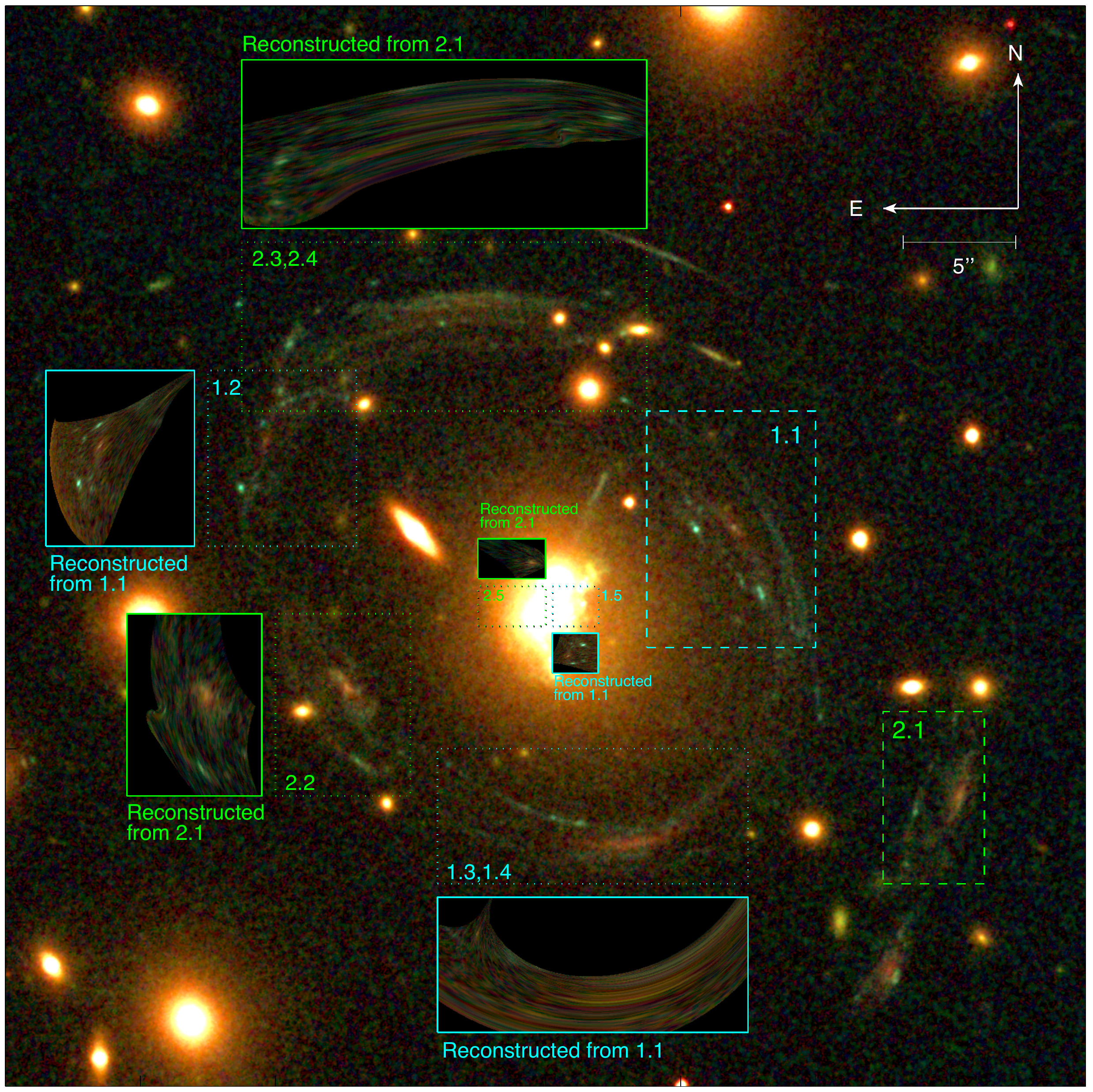}
\caption{Predicted counter images are reconstructed by ray-tracing
  through the best-fit model. We de-lens each pixel in the boxes around image 1.1 
  (cyan, dashed) and image 2.1 (green, dashed) to find its source
  position, and then search the image plane for pixels that are
  lensed to the same source location.  
The resulting model-predicted images of each source are shown, marked in solid lines,
and displaced from their actual locations for clarity. The lensed
positions are marked with dotted lines. 
Images of of source~\#1 and source~\#2 are marked in cyan and green,
respectively.
We find a striking agreement between the model
predictions and the observed image, demonstrating the robustness of
the model and its ability to predict the locations, morphology and
brightness of lensed features.  The color composite is rendered from
F160W, F606W, F390W. 
}
\label{fig.counterimages}
\end{figure*}

\subsection{Strong Lens Model}\label{s.model}
The lens plane is modeled with \texttt{Lenstool} (Jullo et
al. 2007), based on the identification of arcs and galaxies in the new \hst~
imaging, and redshifts from Bayliss et al. (2011). The minimization is done in both the
source plane and in the image plane, using a Markov Chain Monte Carlo
minimizer. The modeling is done iteratively, starting with
the high confidence arcs, and adding on constraints as other multiply
imaged features are identified with the help of the model.
We use as lensing constraints individual emission knots in each of the
lensed images, and the spectroscopic
redshift of source~\#1 (Table~\ref{table.arcs}). The redshifts of the other two sources are
set as free parameters with a lower prior of $z>1.49$, as indicated by the
spectroscopy, while forcing knots that belong to the same galaxy to
have the same redshift.  Figure~\ref{fig.lensmodel} shows the lensed systems,
with the emission knots that were used as constraints labeled.
The lens plane is represented by several pseudo-isothermal ellipsiodal mass distribution 
(PIEMD\footnote{This profile is formally the same as dual Pseudo Isothermal 
Elliptical Mass Distribution (dPIE, see El{\'{\i}}asd{\'o}ttir et al. 2007).}) halos, 
described by the following parameters: position $x$, $y$; a fiducial velocity 
dispersion $\sigma_0$; a core radius $r_{core}$; a truncation radius $r_{cut}$; ellipticity 
$e=(a^2-b^2)/(a^2+b^2)$, where $a$ and $b$ are the semi major and semi 
minor axes, respectively; and a position angle $\theta$. 
Cluster galaxies were selected by their F606W-F814W color in a
color-magnitude diagram, also matching the colors of the spectroscopically-confirmed
cluster members from Bayliss et al. (2011).  All galaxies brighter than F814W 23.5 mag in the
WFC3 field of view were included in the model. 
 Each cluster galaxy is assigned a PIEMD halo,
with positional parameters ($x$, $y$, $e$, $\theta$) 
that follow their observed measurements, $r_{core}$ fixed at 0.15 pc, and 
$r_{cut}$ and $\sigma_0$ scaled with their luminosity (see Limousin et al.\ 2005 
for a description of the scaling relations). 
We allow all the parameters of the cluster PIEMD to vary, except for 
$r_{cut}$, which is not constrainable by the lensing evidence and was thus set to 
1.5 Mpc. We also include a group-size PIEMD halo SE of the BCG, to
represent the dark matter halo of a small group of galaxies. 
We initially allowed all of its parameters
to vary, however, some parameters were not sensitive
to the lensing constraints, and a large range of values within the
priors was allowed. 
We therefore arbitrarily fixed $r_{cut}$, and $r_{core}$. The
positional parameters, $x$ and $y$ are strongly correlated with each
other (following the linear relation $y=\Pbxy x + \Paxy$, where $x$
and $y$ are measured in arcseconds East and North of the BCG, see table~\ref{table.lensmodel}), but are
otherwise only weakly constrained by the lensing 
evidence. 
We have also experimented with adding another group-size halo component
in the direction of the bright galaxy 27'' North of the BCG, which is
the second brightest galaxy in the cluster core, by
allowing the parameters of that galaxy to vary. Similar attempts were
made by freeing the parameters of the central galaxies and other
galaxies close to lensed features.  These iterations did not result in
better models, and thus the parameters of all the cluster galaxies are set
according to the same scaling relations. 

\subsection{Results} 
We find that the cluster is well represented by two cluster- or group-sized halos
plus contribution from galaxy-scale halos. The first halo is centered close to the BCG,
and the second one $\sim56\arcsec$ south-East of the BCG, contributing
shear to the lensing potential. 

Table~\ref{table.lensmodel} lists the best-fit 
parameters and uncertainties, and values of fixed parameters. The model 
uncertainties were determined through the MCMC sampling of the parameter 
space and $1-\sigma$ limits are given. The image plane RMS of the best-fit
lens model is \implanerms \arcsec. 

To demonstrate the predictive power of the model and provide a visual
indication of its robustness, we show in
Figure~\ref{fig.counterimages} image reconstructions of the
counter-images of 1.1 and 2.1. The reconstruction of counter images is computed by
first ray-tracing each pixel in a rectangular region to the source plane
and calculating its source position, and then searching for other pixels
in the image plane that are de-lensed to the same source
position. These pixels are then assigned the same flux values as the
pixel that was ray-traced. The result is a model prediction of the
appearance of all the lensed images, from one
of the lensed images of the source (image 1.1 and 1.2 are selected to be the starting point
of this exercise because they are complete images of their respective sources).
In the figure, the predicted images
are displaced from their predicted location in order to allow a visual
comparison between the model-prediction and the data. 
We find good agreement between the obesrved and reconstructed counter
images of images 1.2,1.3,1.4 and 2.2,2.3,2.4, both in position and
morphology. 
The predicted fifth
image of each system cannot be uniquly identified in the imaging data
(see \S~\ref{s.core}). 

\begin{figure}
\centering
\includegraphics[scale=0.5]{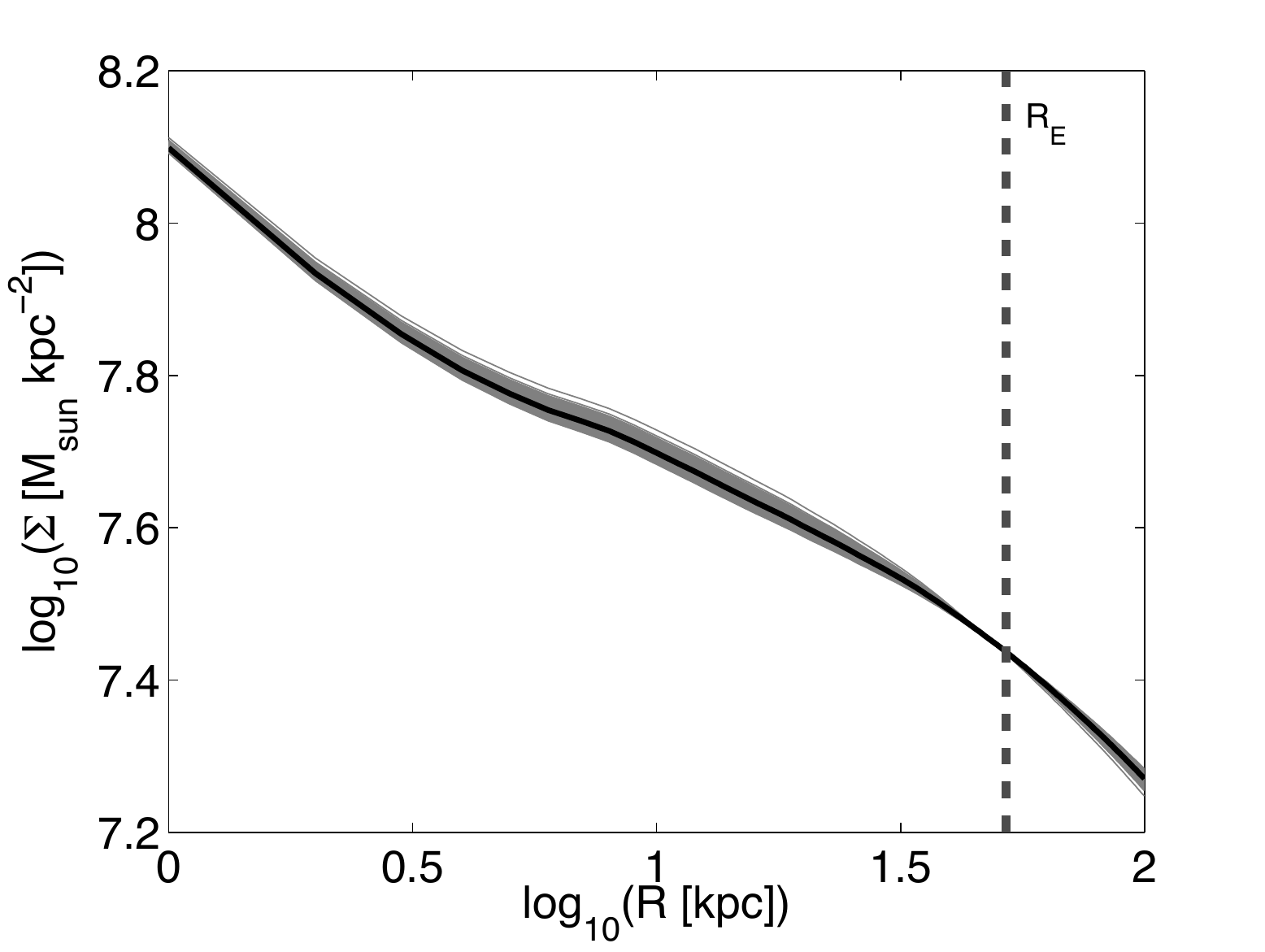}
\caption{{The mass profile of \clustername~ from the
    strong lensing model. The black line indicates the best-fit model,
    and the gray area shows the uncertainty, derived from a suite of
    models with parameters drawn from the MCMC analysis, representing
    a $1-\sigma$ uncertainty in the parameter space. The Einstein
    radius, $R_E=$\thetaE, is indicated by a vertical dashed line. 
}}
\label{fig.massprofile}
\end{figure}

Strong lensing analysis produces a measurement of the mass in the
region of the strong lensing
evidence. We thus refrain from extrapolating the model to larger
radii, and report the  total mass enclosed within $100$ kpc,
approximately the area in which the lensing constraints are found: 
$M(<100~{\rm kpc}) =~$\mass. The projected mass density profile out to
100 kpc is plotted in Figure~\ref{fig.massprofile}.

The PIEMD fiducial velocity dispersions of the
two halos are $\sigma_0$=\Psigma~  km~s$^{-1}$ and
$\sigma_0$=\PsigmaA~  km~s$^{-1}$  for the main and secondary halos,
respectively. To compare the PIEMD parameter to the 
velocity dispersion that was measured by Bayliss et al. (2011), we
need to take into account that the PIEMD fiducial velocity dispersion
is not exactly the velocity dispersion one measures from
the distribution of galaxy redshifts. 
We use the PIEMD formalism given by El{\'{\i}}asd{\'o}ttir et al. (2008), and
integrate their equations (A29) and (A30) numerically to find
the ratio between a measured velocity dispersion and the
PIEMD fiducial velocity dispersion. 
For the best-fit parameters found in our model ($r_{core}/r_{cut}=$\Prcorercut), and the
radius in which the velocity dispersion was measured ($145\arcsec$),
the ratio between a measured velocity dispersion and the
PIEMD fiducial velocity dispersion is
$<\sigma_P>/\sigma_{0,\rm{PIEMD}}=$\sigmacorr, assuming no anisotropy and a
spherically symmetric profile.
The corrected best-fit velocity dispersion of the main halo is thus
$\sigma_{0,\rm{model}} = $\Psigmacorr~  km s$^{-1}$. This value is formally 
in agreement (within uncertainties) with the velocity dispersion that was measured by Bayliss
et al. (2011), $\sigma_{0,\rm{obs}} = 998^{+120}_{-194}$ km s$^{-1}$
from 11 galaxies; nevertheless, the offset can be a result of the
complexity of the mass distribution, and the ellipticity of the halos.   

Since the lens model is composed of two elliptical halos the
velocity dispersion of the system as a whole will account for the sum
of their masses. 
For this comparison, we integrate the enclosed
projected mass density of
a single spherical PIEMD halo out to $100$ kpc and find that in order
to produce the model-derived enclosed mass the corrected PIEMD fiducial velocity
dispersion would be 892 km s$^{-1}$, closer to the observed value.

The best fit model predicts a redshift of \PredshiftB~ for source~\#2, and
\PredshiftC~ for source 3, with a strong correlation between the
redshift of source~\#2 and the main halo parameters. The relation
between the redshift of source~\#2 and the velocity dispersion is consistent with
$\sigma_0 = (\Pazsig \Pbzsig \times z)$ km s$^{-1}$. A similar linear correlation is found with the core
radius and with the mass of the halo, due to the analytical relation between the core radius,
velocity dispersion, and mass in the PIEMD profile. Similarly, the
projected mass density enclosed
in a radius of $100$ kpc is correlated with the redshift of
source~\#2, and follows the relation $M(<100 ~{\rm kpc}) = (\Pazmass
\Pbzmass \times z )\times 10^{12}$\msun.
We thus expect to derive higher values of $\sigma_0$, $r_{core}$, and
enclosed mass if a higher redshift is measured for source~\#2 in future spectroscopic
observations. 

\begin{figure*}
\centering
\includegraphics[scale=0.4]{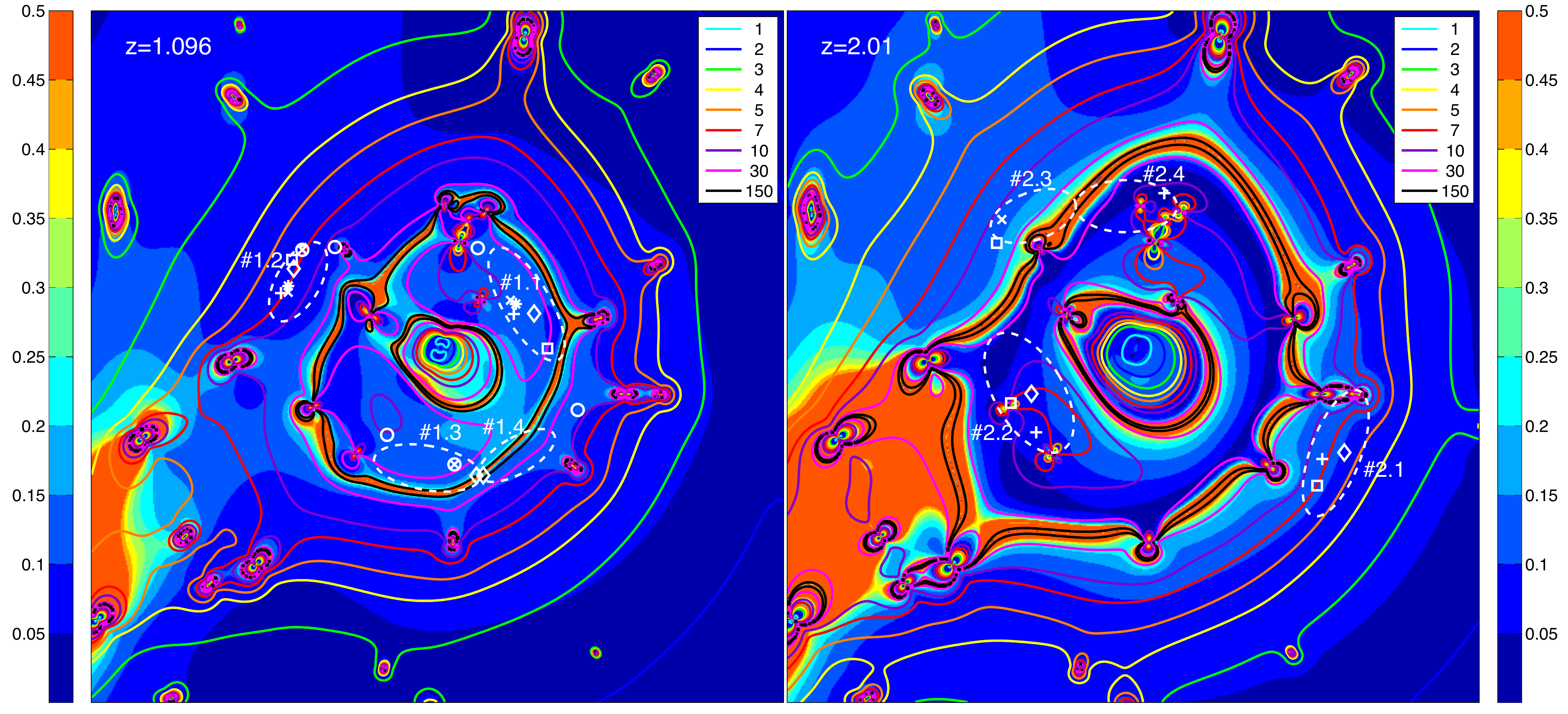}
\caption{{The magnification and its uncertainty for source~\#1 ({\it left}) and
    source~\#2 ({\it right}).  The contours indicate the value of the
    magnification, and the color gradient indicates the uncertainty in
    the magnification in each position. The uncertainty is given in
    units of $\Delta\mu/\mu$.  The field of view and scale are
    the same as those in Figure~\ref{fig.lensmodel}.  To guide the
    eye, we plot the positions of the images of sources \#1 and \#2,
    and the clumps that were used as constraints in the model. 
}}
\label{fig.magnific}
\end{figure*}

We derive a best-fit lensing magnification and uncertainties from a suite of
models, computed 
from sets of parameters taken from the MCMC sampling, and representing 
1-$\sigma$ in the parameter space.
Figure~\ref{fig.magnific} shows the magnification of the best-fit
model as contours, and its
uncertainty as color scale, in each position in the image plane
for two source redshifts: a source at $z=$\zarcA~ and a source at
$z=2.01$, which is the predicted
redshift of source~\#2.  The uncertainties 
take into account the distribution of model-predicted
redshifts for source~\#2. We note that the relative uncertainty
($\Delta\mu/\mu$) is largest close to the critical curves, as well as
in the direction of the secondary halo, whose position is poorly
constrained. Nevertheless, in the locations of the lensing
constraints, the magnification uncertainty is typically below 20\%,
with the exception of arcs very close to the critical curves. 

The Einstein radius, $R_E$, is often used as an indicator of the lensing cross
section. In a spherically symmetric lens the tangential critical curve
is a circle, and $R_E$ would be the radius of that circle. 
Here, we define $R_E=\sqrt{A/\pi}$, where $A$ is the
area enclosed by the tangential critical curve; for a circular lens
the two definitions are identical.  The lensing configuration in this
cluster can provide a robust estimate of $R_E$. 
We find an Einstein radius of ${\rm R}_E=$\thetaE~ for a source at
$z=$\zarcA~ and  ${\rm R}_E=$\thetaEB~ for a source at the model
predicted redshift of source~\#2. The uncertainty of the latter takes
into account the distribution of predicted redshifts for  source~\#2.

\begin{deluxetable*}{lccccccc} 
 \tablecolumns{8} 
\tablecaption{Best-fit lens model parameters  \label{table.lensmodel}} 
\tablehead{\colhead{Halo }   & 
            \colhead{RA}     & 
            \colhead{Dec}    & 
            \colhead{$e$}    & 
            \colhead{$\theta$}       & 
            \colhead{$r_{\rm core}$} &  
            \colhead{$r_{\rm cut}$}  &  
            \colhead{$\sigma_0$}\\ 
            \colhead{(PIEMD)}   & 
            \colhead{($\arcsec$)}     & 
            \colhead{($\arcsec$)}     & 
            \colhead{}    & 
            \colhead{(deg)}       & 
            \colhead{(kpc)} &  
            \colhead{(kpc)}  &  
            \colhead{(km s$^{-1}$)}  } 
\startdata 
Halo 1    & \Px        & \Py           & \Pe       & \Ptheta        &\Prc     & [1500]    & \Psigma  \\ 
%BCG         & [0.0]       & [0.0]        & [0.067] & [-81.3]          & [0.19] & [38.2]     & \PsigmaA  \\ 
Halo 2        & \PxA       & \PyA        & \PeA & \PthetaA          & [50] & [1500]     & \PsigmaA  \\ 
%\cutinhead{Fixed components}
%Halo2       & [7.11]     & [290]        & [0.0] & [0.0]          & [100] & [1500]     & [950]  \\ 
L* galaxy  & \nodata & \nodata & \nodata & \nodata &  [0.15]  &     [40]&  [170]  \\

\enddata 
 \tablecomments{All coordinates are measured in arcseconds East and
   North of the center of the BCG, at [RA, Dec]=[232.79429 34.240312]. The
   ellipticity is expressed as $e=(a^2-b^2)/(a^2+b^2)$. $\theta$ is
   measured North of West. Error bars correspond to 1-$\sigma$
   confidence level as inferred from the MCMC optimization. Values in
   square brackets are for parameters that were not optimized. The
   location and the ellipticity of the matter clumps associated with
   the cluster galaxies and the BCG were kept fixed according to their
   light distribution, and the fixed parameters determined through
   scaling relations (see text). }
\end{deluxetable*}

\subsubsection{Comparison to Previous Work}\label{s.comp}
Gralla et al. (2011) presented a joint analysis of SZ and strong
lensing in 10 strong lensing clusters, including \clustername. 
The Einstein radius they report for \clustername~ is $R_E=12.3$\arcsec~
for a source at $z=1.096$,
with an estimated {\it statistical} uncertainty of up to 5\%. This
measurement is higher than the one we find here, $R_E=$\thetaE, for a
source at $z=\zarcA$. The reason for the discrepancy is that the  
lensing analysis in Gralla et al. (2011) was based on ground-based
imaging: in those data, the interpretation of the lensing
configuration was ambiguous and the identification of counter images
was lacking.  

A strong lensing model for \clustername~ was also published by
Oguri et al. (2012), based on identification of the images
of source~\#1 in Subaru data, and on its spectroscopic redshift from
Bayliss et al. (2011). Oguri et al. (2012) report
an Einstein radius of $11.7\pm1.2 \arcsec$ for a source at $z=1.096$,
in agreement with our result (within uncertainties).

\begin{figure}
\centering
\includegraphics[scale=0.22]{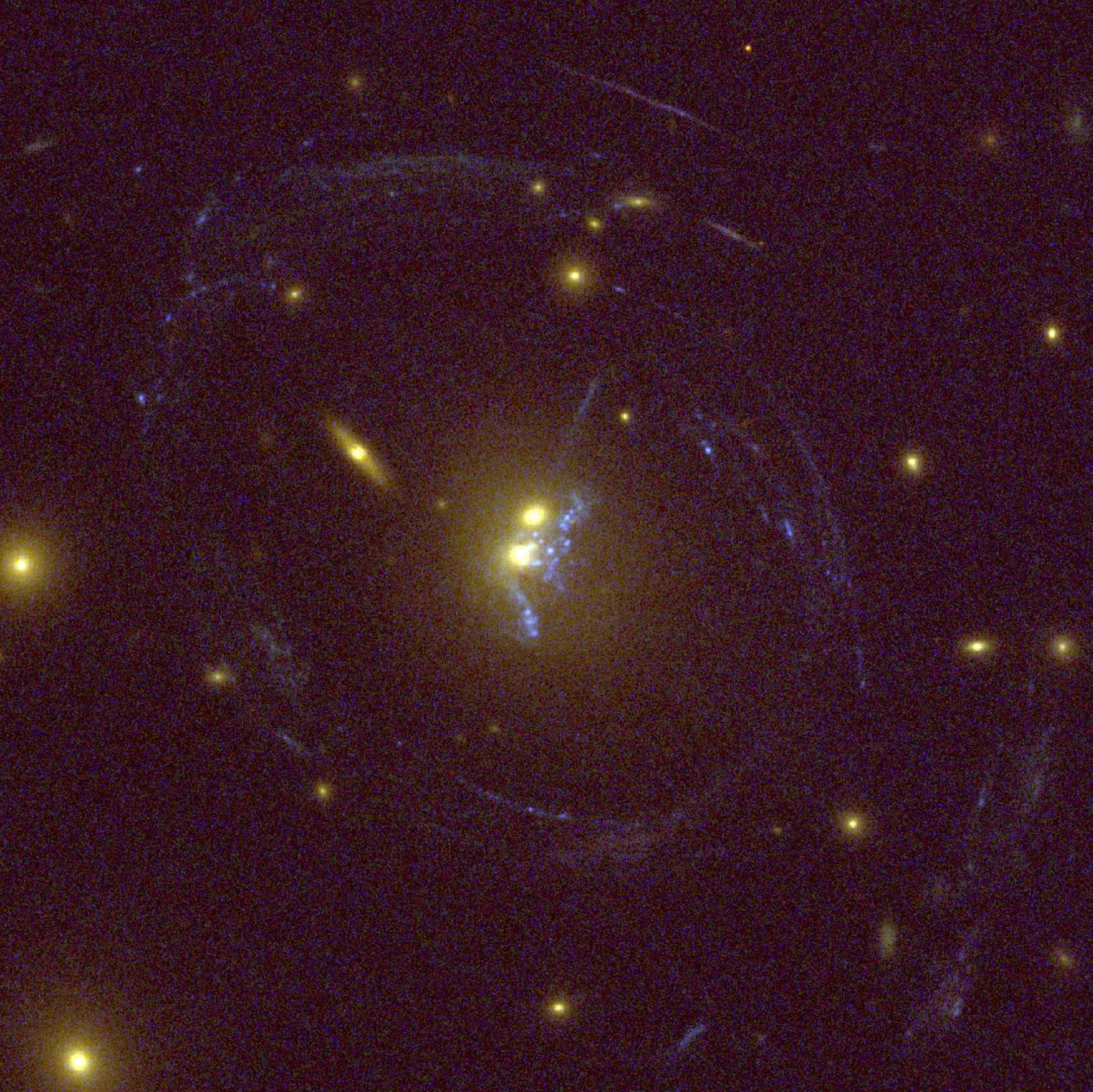}
\caption{{WFC3/UVIS color composite image of a 40'' region
around the core of \clustername~: F814W (red) ,F606W (green), F390W
(blue). High color scale values were selected in order to reveal the resolved star forming
emission knots at the center of the cluster, within 3'' from the two 
central galaxies (Tremblay et al. 2014).  Note that these emission knots are
significantly brighter than any of the emission knots in the lensed
images of the background galaxies. If these emission knots were coming
from  lensed images of the background galaxies, they would have been
less magnified and appear fainter than the tangential images.     
}}
\label{fig.core}
\end{figure}

\begin{figure*}
\centering
\includegraphics[scale=0.47]{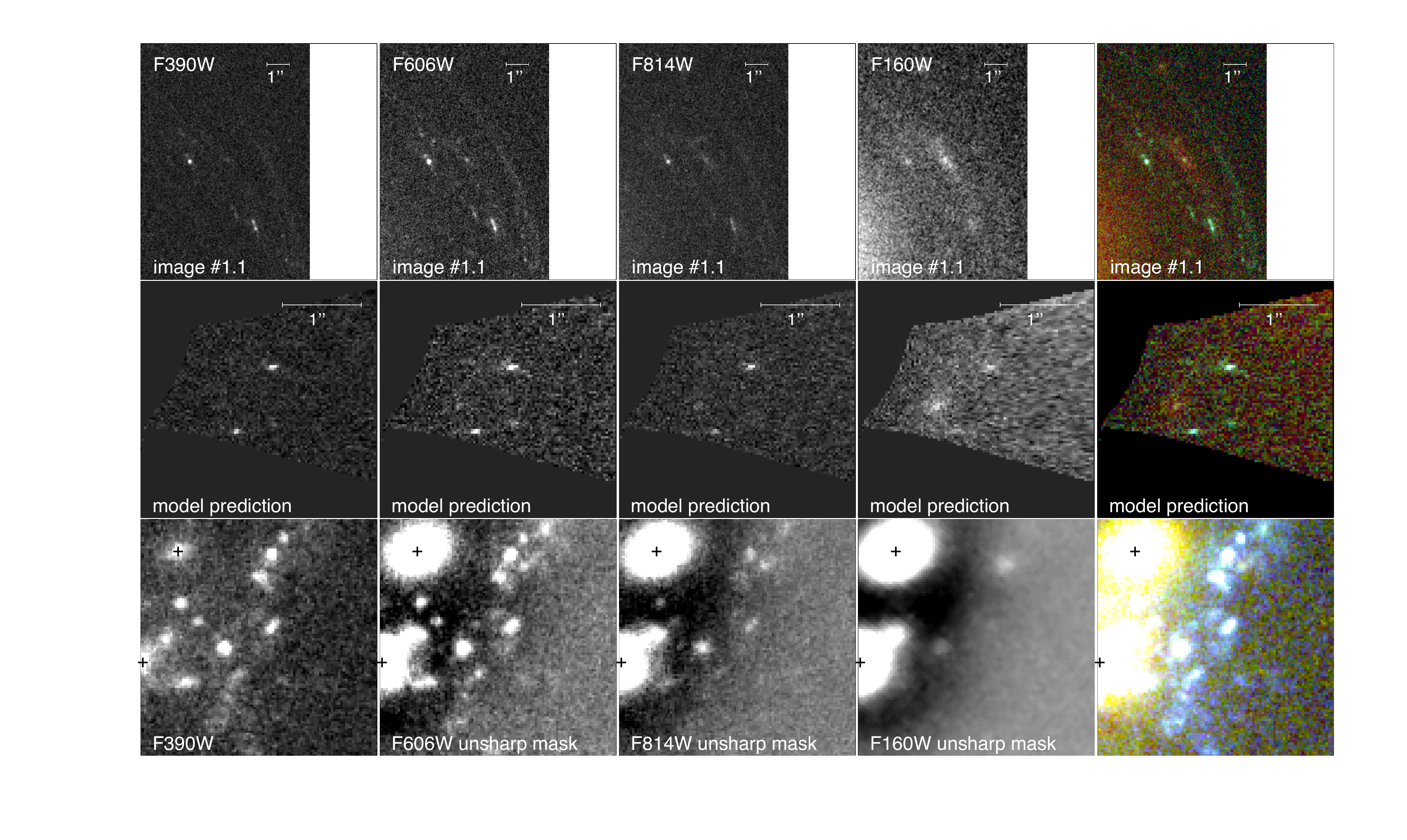}
\caption{{Model prediction of a fifth image of source~\#1,
    near the central cluster galaxies. The top row shows image 1.1
    (see Figure~\ref{fig.lensmodel}) in each of the \hst~ bands, and in
    a composite color image. 
    The predictions, shown in the
    middle row, are computed by
    delensing and relensing the area around image \#1.1 (the same area
    that is shown in the top row) through the best-fit
    lens model. The bottom row shows the observed images, at the same 
    location in the image plane as in the middle
    row, showing that the predicted images cannot be easily detected
    in current imaging depth. To reveal the star forming clumps that
    are buried in the 
    light of the central galaxies, we show in the bottom row an
    unsharp mask for the 
    redder bands. A horizontal line indicates a $1\arcsec$ scale in
    each panel, and the black cross indicates the two cluster galaxies
    at the core of \clustername. To guide the eye, we show composite
    color renditions in the 
    rightmost column: F160W, F606W, F390W in the top and middle
    panels, and F814W, F606W, F390W in the bottom panel.  This figure
    demonstrates that the predicted central 
    image of source 1 is significantly fainter than the observed 
    emission clumps at the center of the cluster, indicating that this
    emission is not due to light from the background lensed galaxy.
}}
\label{fig.central_prediction}
\end{figure*}

\section{Discussion: The Core of \clustername}\label{s.core}
The high resolution, multiband shallow {\it HST} images of
\clustername~ reveal unusual star formation activity near the core
of the cluster (Figure~\ref{fig.core}). This star formation is detected in the form
of multiple blue emission knots on a scale of approximately 0.5-1 kpc
in diameter,
and spanning $\sim$27 kpc across. 
 In a companion paper (Tremblay et al. 2014), we analyze this emission and report on the star
 forming activity and the spatial distribution of the star forming regions,
 and discuss limits on their age and formation scenarios.  
In this paper, we determine what portion of the observed emission
is attributed to images of a background galaxy or galaxies. 

As described above, the lens model predicts the formation of a fifth
lensed image for both source~\#1 and source~\#2, and a third image for
source~\#3.
These images are predicted to lie close to the center of the cluster,
and are often referred to as central images. In most lensing
configurations, the central image is not highly magnified, and often 
it is demagnified (i.e., $\mu<1$) and undetectable. In the case of \clustername, the lens model predicts
that the central image of source~\#1 is somewhat magnified, by approximately 2, with
a spatial size comparable to that of the SF region. Nevertheless, it
would be significantly less magnified than images
1.1-1.4. Quantitatively, the brightest emission knot in the predicted fifth image of source~\#1 would
be at least two magnitudes fainter than that in image 1.2 (marked with
a cyan cross in Figure~\ref{fig.lensmodel}), which would result
in F390W magnitude $>27$ mag. The emission knots that are detected in
the F390W image, in the innermost $3\arcsec$, are significantly
brighter, with magnitudes ranging from 23 to 26 mag. 
The central images of sources~\#2 and \#3 are predicted to be demagnified, and
are fainter than our detection limit.
 Figure~\ref{fig.central_prediction} shows the predicted central image
 of source~\#1 in
the four \hst~ bands, and the corresponding area in the images. 
Since the light of the central cluster galaxies dominate the center of
the cluster in all but the bluest band, we created an unsharp mask to
reveal the central morphology. The unsharp mask was created by
convolving the image with a gaussian with a width of 20 pixels and subtracting
it from the original image.  We note that the \hst~ point spread function was not
applied to the predicted image in
Figure~\ref{fig.central_prediction}. 
The predicted image has two faint knots in the F390W band; a few more
faint knots in the F606W bands, only one knot in the F814W
band; and in the F160W, the brightest feature is the center of the lensed
galaxy. 
 It is thus possible that at most one or two of the knots are brighter
 than the detection limit in
the F390W band, but they do not correspond to the much brighter knots
that are reported on in Tremblay et al. (2014). The center of the
lensed galaxy, which is most noticeable in the F160W image, does not
have a counterpart at the predicted location within the detection
limit and colors. It is possible that the faint fifth image is
obscured by the light of the much brighter star formation knots.  

Further evidence against the identification of the bright emission
knots as lensed galaxies comes from presence of H$\alpha$ flux at the
cluster redshift, in spectroscopic observations using
the Nordic Optical Telescope (NOT). These observations are described
in Tremblay at al. (2014). Had these emission knots been a counter
image of source~\#1, these observations would have revealed strong
 [O II]$\lambda\lambda3727$ emission at $z=$\zarcA. 
A careful reduction of the NOT spectra and a thorough examination of
the data reveals no trace of emission from the fifth image of
source~\#1 where this emission line is expected.

We conclude the emission in the clumps that are
detected close to the two central galaxies is too bright to be coming from counter
images of the lensed galaxies, and rule out that a faint central
image of a background source contributes significantly to this emission. 

In summary, we present new \hst~ imaging data and a lens model of the
strong lensing cluster \clustername. We detect three lensed background
sources, and uniquely identify their multiple images that stretch to
form giant arcs around the core of the
cluster.  We find that the lens model is best represented by
two cluster-scale halos, with contributions from cluster-member
galaxies. The mass enclosed in the innermost 100 kpc is
\mass, in agreement with previous estimates. The lens model predicts
the formation of a faint image close to the center of the cluster. At
this position, we detect bright emission knots, most prominent in the
bluer \hst~ filters. Based on the lens model presented here, we rule
out the possibility that this emission is coming from a counter image
of one of the giant arcs, and support the interpretation of Tremblay et
al. (2014) that this emission is due to star formation at the core of
\clustername.

\acknowledgments
Support for program number GO-13003 was provided by NASA through a
grant from the Space Telescope Science Institute, which is operated by
the Association of Universities for Research in Astronomy, Inc., under
NASA contract NAS5-26555. We also present results based on
observations with the Nordic Optical Telescope, operated by the Nordic
Optical Telescope Scientific Association at the Observatorio 
del Roque de los Muchachos, La Palma, Spain, of the Instituto de
Astrofisica de Canarias.

\end{document}